\documentstyle[12pt,fleqn]{article}
\newcommand{\be}{\begin{equation}}
\newcommand{\ee}{\end{equation}}
\newcommand{\ssr}[1]{{\, \strut ^{w}}\! {#1}}
\newcommand{\ssd}[1]{{\, \strut ^{d}}\! {#1}}
\newcommand{\ssm}[1]{{\, \strut ^{m}}\! {#1}}

\begin{document}
\title{\bf Asymptotic Conformal Yano--Killing Tensors for Schwarzschild
       Metric}
\author{Jacek Jezierski\\
Department of Mathematical Methods in
Physics, \\ University of Warsaw,
ul. Ho\.za 74, 00-682 Warsaw,
Poland}
\maketitle

\begin{abstract}
The {\em asymptotic conformal Yano--Killing tensor} proposed in
\cite{JJ-s2} is analyzed for Schwarzschild metric and
tensor equations defining this object are given.
The result shows that the Schwarzschild metric (and other
metrics which are asymptotically ``Schwarzschildean'' up to $O(1/r^2)$ at
spatial infinity) is among the metrics fullfilling stronger asymptotic
conditions and supertranslations ambiguities disappear.
It is also clear from the result that 14 asymptotic gravitational charges
are well defined on the ``Schwarzschildean'' background.
\end{abstract}

\section{Introduction}
We have proposed in \cite{JJ-s2} the charged solutions of spin-2 equations.
The new charges result in a natural way from a
geometric formulation of the ``Gauss law'' for the gravitational
charges, defined in terms of the Riemann tensor (equation \ref{*R*}).
  It leads to the notion of the {\em conformal Yano--Killing tensor}.
A conformal Yano--Killing (CYK) equation (\ref{CYK}) posseses
twenty--dimensional space of
solutions for flat Minkowski metric in four--dimensional spacetime.
This can be easily seen from our analysis when we pass to the limit with
mass parameter $m$ ($m\rightarrow 0$).

A natural application of the construction of CYK tensor to the description of
asymptotically flat spacetimes was proposed in \cite{JJ-s2}. It
allows us to
define an asymptotic charge at spatial infinity without supertranslation
ambiguities. The existence or nonexistence of the
corresponding asymptotic CYK tensors can be  chosen as a criterion for
classification of asymptotically flat spacetimes.
We show in this article that Schwarzschild metric is an example of a nice
asymptotically flat spacetime from this point of view. It possesses a full
set of 14 asymptotic CYK tensors.

\section{Conformal Yano--Killing tensors and asymptotically flat
spacetimes}

Let $Q_{\mu\nu}$ be an antisymmetric tensor field
fulfilling the following condition introduced by Penrose
(see \cite{Pen-Rin} and \cite{JNG}):
\begin{equation}\label{Q}
Q_{\lambda (\kappa ;\sigma)} -Q_{\kappa (\lambda ;\sigma)} +
\eta_{\sigma[\lambda} Q_{\kappa ]}{^\delta}_{;\delta} =0
\end{equation}

It is easy to check that equation (\ref{Q}) is
equivalent to the following one:
\begin{equation}\label{CYK}
Q_{\lambda \kappa ;\sigma} +Q_{\sigma \kappa ;\lambda} -
\frac{2}{3} \left( g_{\sigma \lambda}Q^{\nu}{_{\kappa ;\nu}} +
g_{\kappa (\lambda } Q_{\sigma)}{^{\mu}}{_{ ;\mu}} \right) =0
\end{equation}
The tensor fulfilling the equation (\ref{Q}) or (\ref{CYK}) we proposed
to call the {\em conformal Yano Killing tensor} (or simply CYK). The CYK
tensor is a natural ``conformal'' generalization of the Yano tensor.

Consider an asymptotically flat spacetime (at spatial infinity),
fulfilling the Einstein equations. Suppose,
moreover, that the energy--momentum tensor of the matter vanishes
around spatial infinity (``sources of compact support'').
Let us analyze, for simplicity,
this situation in terms of an asymptotically flat coordinate system.
We suppose that there exists an (asymptotically
Minkowskian) coordinate system $(x^\mu)$:
\[ g_{\mu\nu} - \eta_{\mu\nu} =O( r^{-b}) \; \; \; \; \;
 g_{\mu\nu,\lambda} =O( r^{-b -1}) \] \noindent
where $\displaystyle r:= \sum_{k=1}^{3} (x^k)^2$ and typically $b=1$ (but
$1\geq b > \frac 12$ is also possible).

For a general asymptotically flat metric we cannot expect that the
equations (\ref{Q}) and
(\ref{CYK}) admit any solution. Instead, we assume that the
left--hand side of (\ref{CYK}):
 \be \label{Q3}
 {\cal Q}_{\lambda\kappa\sigma}:= Q_{\lambda \kappa ;\sigma} +Q_{\sigma
\kappa ;\lambda} -
\frac{2}{3} \left( g_{\sigma \lambda}Q^{\nu}{_{\kappa ;\nu}} +
g_{\kappa (\lambda } Q_{\sigma)}{^{\mu}}{_{ ;\mu}} \right)
   \ee \noindent
has certain asymptotic behaviour at spatial infinity
\begin{equation}\label{QAF}
 {\cal Q}_{\mu\nu\lambda} =O( r^{-c})  \end{equation}
 On the other hand, suppose that  $Q_{\mu\nu}$ behaves asymptotically as
follows:
 \[ Q_{\mu\nu} =O( r^{a}) \; \; \; \; \;
 Q_{\lambda \kappa , \sigma} =O( r^{a-1})  \]
\noindent
Moreover, suppose that the Riemann tensor $R_{\mu\nu\kappa\lambda}$
behaves asymptotically as follows:
\[ R_{\mu\nu\kappa\lambda} =O( r^{-b -1 -d}) \] \noindent
It can be easily checked (see e.g. \cite{JNG}) that the vacuum Einstein
equations imply the following equality:
 \begin{equation}\label{*R*}
 \nabla_{\lambda} \left( {^*\!}R{^*}^{\mu\lambda}{_{\alpha\beta}}
Q^{\alpha\beta} \right) =  \frac 13
R^{\mu\lambda \alpha\beta} Q_{\alpha\beta\lambda}
 \end{equation}

The left--hand side of (\ref{*R*}) defines an asymptotic charge provided
that the right--hand side vanishes sufficiently fast at infinity. It is easy to
check that, for this purpose, the exponents $b,c,d$ have to fulfill the
inequality $ b + c + d > 2 $.
In typical
situation when $b=d=1$, the above inequality simply means that $c > 0$. In
this case a weaker condition is also possible (for example ${\cal
Q}_{\mu\nu\lambda} =O( (\ln r)^{-1-\epsilon})$ with $\epsilon >0$).
Moreover, when ${\cal Q}_{\mu\nu\lambda}$ vanishes the formula (\ref{*R*})
gives ``pure'' charge (not only asymptotic).

Let us define an {\em asymptotic conformal Yano--Killing tensor} (ACYK) as an
antisymmetric
tensor $Q_{\mu\nu}$ such that ${\cal Q}_{\mu\nu\lambda} \rightarrow
0$ at spatial infinity.
For constructing the ACYK tensor we can begin with the solutions of (\ref{Q})
in flat Minkowski space. Asymptotic behaviour at infinity of these flat
solutions explain why we expect for any ACYK tensor the following
behaviour:
\[ Q_{\mu\nu} = {^{(2)}\!}Q_{\mu\nu} + {^{(1)}\!}Q_{\mu\nu} +
{^{(0)}\!}Q_{\mu\nu} \] \noindent
where ${^{(2)}\!}Q_{\mu\nu} =O( r^2)$, ${^{(1)}\!}Q_{\mu\nu} =O( r)$ and
${^{(0)}\!}Q_{\mu\nu} =O( r^{1-c})$.

It is easy to verify that $c \geq b +1 -a$ and if $b=1$
than for $a=2$ we have $c \geq 0$. This means that in a general situation
there are no solutions of (\ref{QAF}) with nontrivial
${^{(2)}\!}Q_{\mu\nu}$ and $c>0$.
This is the origin of the difficulties with the definition of the angular
momentum. On the other hand it is easy to check that the energy--momentum
four--vector and the dual one are well defined ($a=c=1$)  and the condition
$b + d > 1$ can be easily fulfilled (typically $b=d=1$).

We proposed in \cite{JJ-s2} a new, stronger definition of the asymptotic
flatness.  The definition is motivated by the above discussion.

Suppose that there exists a coordinate system $(x^\mu)$
such that:
\[ g_{\mu\nu} - \eta_{\mu\nu} =O( r^{-1}) \]
\[ \Gamma^{\kappa}{_{\mu\nu}} =O( r^{-2}) \]
\[ R_{\mu\nu\kappa\lambda} =O( r^{-3}) \] \noindent
In the space of ACYK tensors fulfilling the asymptotic condition
\begin{equation}\label{Q1}
Q_{\lambda \kappa ;\sigma} +Q_{\sigma
\kappa ;\lambda} -
\frac{2}{3} \left( g_{\sigma \lambda}Q^{\nu}{_{\kappa ;\nu}} +
g_{\kappa (\lambda } Q_{\sigma)}{^{\mu}}{_{ ;\mu}} \right) =
{\cal Q}_{\lambda\kappa\sigma} =O( r^{-1})
  \end{equation}
we define the following equivalence relation:
\begin{equation}\label{rel} Q_{\mu\nu} \equiv Q_{\mu\nu}'
\Longleftrightarrow Q_{\mu\nu}  - Q_{\mu\nu}' =  O(1)
 \end{equation}
for $r \rightarrow \infty$.
We assume that the space of equivalence classes defined by (\ref{Q1}) and
(\ref{rel}) has a finite dimension $D$ as a vector space.
The maximal dimension $D = 14$ correspond to the situation where there
are no supertranslation problems
in the definition of an angular momentum.
In the case of spacetimes for which $D<14$
the lack of certain ACYK tensor means that the corresponding charge
is not well defined.

\section{Conformal Yano--Killing tensors for Schwarzschild metric}

For our purposes we need to specify the formula (\ref{CYK})
to the  special case of the Schwarzschild metric $g_{\mu\nu}$:
\be \label{Sch}
g_{\mu\nu}\, dx^\mu \, dx^\nu\, =
-\left(1-\frac{2m}r \right) dt^2 + \left(1-\frac{2m}r\right)^{-1} dr^2
 +r^2 d\theta^2 + r^2 \sin^2 \theta d\varphi^2
\ee
We use radial coordinates: $x^3=r$,
$x^1=\theta $, $x^2=\varphi$. Moreover, $t=x^0$ denotes the time
coordinate. We consider only part of the Schwarzschild spacetime far away
from the horizon, $r \gg m$.

We use the following convention for indices: greek indices $\mu, \nu,
\ldots$ run from 0 to 3;
$k,l, \ldots$ are spatial coordinates and run from 1 to 3;
$A,B,\ldots$ are spherical
angles $(\theta, \varphi)$ on a two-dimensional sphere $S(r):= \{
r=x^3=const. \}$ and run from 1 to 2.
Moreover, let $\eta_{AB}$ denotes two-dimensional metric on $S(r)$.

Let $\displaystyle v:=1-\frac{2m}r$.
There are following non-vanishing Christoffel symbols for the metric
(\ref{Sch}):
\[ \Gamma^{3}{_{AB}}= -\frac vr  \eta_{AB} \, ; \;\;
 \Gamma^{A}{_{3B}}= \frac 1r \delta^A{_B} \, ; \;\;
 \Gamma^{3}{_{00}}= \frac{mv}{r^2}  \, ; \;\;
 \Gamma^{0}{_{30}}= \frac{m}{vr^2}  \, ; \;\;  \Gamma^{A}{_{BC}} \]
\noindent
where $\delta^A{_B}$ is a Kronecker's symbol
 and $\displaystyle \Gamma^{A}{_{BC}}$ are the same as for standard
unit sphere $S(1)$.

For Schwarzschild metric (\ref{Sch}) we obtain the following components of
the equation (\ref{Q3}):
\be\label{003}
{\cal Q}_{003}=Q_{30;0}-\frac 13 g_{00}Q_3{^\nu}{_{;\nu}}=\frac 23
\dot{Q}_{30} + \frac v3  Q_{3A||B}\eta^{AB}
\ee

\be\label{00A}
{\cal Q}_{00A}=Q_{A0;0}-\frac 13 g_{00}Q_A{^\nu}{_{;\nu}}=\frac 23
\dot{Q}_{A0} +  \frac{2mv}{3r^2}  Q_{3A} +
 \frac v3  Q_{AB||C}\eta^{BC} + \frac{v^2}3 Q_{A3,3}
\ee

\be\label{033}
{\cal Q}_{033}=Q_{03;3}-\frac 13 g_{33}Q_0{^\nu}{_{;\nu}}=\frac 23
{Q}_{03,3} -  \frac{2m}{3vr^2} Q_{03} -  \frac{2}{3r} Q_{03}
- \frac1{3v} Q_{0A||B}\eta^{AB}
\ee

\be\label{03A}
{\cal Q}_{03A}=Q_{03;A}+Q_{A3;0}=
{Q}_{03,A} - \frac{1}{r} Q_{0A} +\dot{Q}_{A3} + \frac m{vr^2} Q_{0A}
\ee

\[
{\cal Q}_{0AB}=Q_{0A;B}+Q_{BA;0}-\frac13\eta_{AB}Q_0{^\nu}{_{;\nu}}=
{Q}_{0A||B}- \frac{1}{3}\eta^{CD} Q_{0C||D} \eta_{AB} -\dot{Q}_{AB}+
\]
\be\label{0AB}
 +\frac13 \eta_{AB} \left( \frac{v}{r} Q_{03}  + \frac m{r^2} Q_{03}
 -v Q_{03,3} \right)
\ee

\be\label{30A}
{\cal Q}_{30A}=Q_{30;A}+Q_{A0;3}=
{Q}_{30,A} + \frac{2}{r} Q_{0A} +{Q}_{A0,3} + \frac m{vr^2} Q_{0A}
\ee

\[
{\cal Q}_{A0B}=Q_{A0;B}+Q_{B0;A}+\frac23\eta_{AB}Q_0{^\nu}{_{;\nu}}=
{Q}_{A0||B}+{Q}_{B0||A}+ \frac{2}{3}\eta^{CD} Q_{0C||D} \eta_{AB} +\]
\be\label{A0B}
 -\frac23 \eta_{AB} \left( \frac{v}{r} Q_{03}  + \frac m{r^2} Q_{03}
 -v Q_{03,3} \right)
\ee

\[
{\cal Q}_{33A}=Q_{A3;3}-\frac 13 g_{33}Q_A{^\nu}{_{;\nu}}=\frac 23
{Q}_{A3,3} + \frac{m}{3vr^2} Q_{3A} + \frac{1}{r} Q_{3A} -\frac{1}{3v^2}
\dot{Q}_{0A} +\]
\be\label{33A}
- \frac1{3v} Q_A{^B}{_{||B}}
\ee

\[
{\cal Q}_{3AB}=Q_{3A;B}+Q_{BA;3}-\frac13\eta_{AB}Q_3{^\nu}{_{;\nu}}=
{Q}_{3A||B}- \frac{1}{3}\eta_{AB}\left( Q_3{^C}{_{||C}}
+v^{-1}\dot{Q}_{03} \right)+\]
\be\label{3AB}
 +{Q}_{BA,3}+\frac3r Q_{AB}
\ee

\[
{\cal Q}_{A3B}=Q_{A3;B}+Q_{B3;A}+\frac23\eta_{AB}Q_3{^\nu}{_{;\nu}}=
{Q}_{A3||B}+{Q}_{B3||A}+ \frac{2}{3}\eta_{AB}  Q_3{^C}{_{||C}}+\]
\be\label{A3B}
 + \frac{2}{3v}\eta_{AB} \dot{Q}_{03}
\ee

\[
{\cal Q}_{ABC}=Q_{AB;C}+Q_{CB;A}-\frac23 \left(\eta_{AC}Q^\nu{_{B;\nu}}+
\eta_{B(A}Q_{C)}{^\nu}{_{;\nu}}\right)=
{Q}_{AB||C}+{Q}_{CB||A} +\]
\[
+ \frac{2}{3}\eta_{AC}  Q_B{^D}{_{||D}} - \frac{2}{3}\eta_{B(A}
Q_{C)}{^D}{_{||D}} +\frac{2v}{r}\left( \eta_{AC} {Q}_{3B} +\eta_{B(A}
{Q}_{C)3} \right)
 + \frac{2}{3v} \eta_{AC} \dot{Q}_{0B} + \]
 \be\label{ABC}
 + \frac{2}{3v}\eta_{B(A} \dot{Q}_{C)0}
 + \frac{2}{3}\eta_{AC}\left( v {Q}_{B3,3}+\frac{m}{r^2}Q_{B3} \right)
 -\frac{2}{3}\eta_{B(A} \left( v Q_{C)3,3}+\frac{m}{r^2}Q_{C)3}
\right)
\ee
\noindent
where ``$;$'' denotes four--dimensional covariant derivative with respect
to the Schwarzschild metric $g_{\mu \nu}$, by dot we have denoted as usual
time derivative and symbol ``$||$'' denotes
two-dimensional covariant derivative with respect to the two-metric
$\eta_{AB}$.

{}From (\ref{A0B}) and (\ref{033}) we get:
\be\label{0Aw}
{\cal Q}_{A0B} -v{\cal Q}_{033}=
 Q_{0C||D}\eta^{CD}\eta_{AB}-{Q}_{0A||B}-{Q}_{0B||A}
\ee
Similarly from (\ref{A3B}) and (\ref{003}) we have
\be\label{3Aw}
{\cal Q}_{A3B} +v^{-1}{\cal Q}_{003}=
 Q_{3C||D}\eta^{CD}\eta_{AB}-{Q}_{3A||B}-{Q}_{3B||A}
\ee
Let ${\bf a}$ denotes the two--dimensional Laplace--Beltrami operator on a
unit sphere $S(1)$ and on each sphere $S(r)$ we denote
by $\varepsilon ^{AB}$ the Levi--Civita
antisymmetric tensor such that $r^2\sin\theta \, \varepsilon ^{12} =1$.
For a function $f$ on $S(r)$  we denote by $\ssm f$  its monopole
part, by $\ssd f$ its dipole part and we denote by $\ssr f$ the
remainder, which will be called the ``radiation part''.
The equations (\ref{0Aw}) and (\ref{3Aw}) show that only mono--dipole
part of $Q_{0A}$ and $Q_{3A}$ can have nontrivial asymptotic behaviour,
the higher poles have to vanish at spatial infinity. More precisely:
\[
{\ssr Q}_{0A||B}\eta^{AB}=O\left(\frac 1r \right) \, , \;\;\;\;
{\ssr Q}_{0A||B}\varepsilon^{AB}=O\left(\frac 1r \right) \]
\[
{\ssr Q}_{3A||B}\eta^{AB}=O\left(\frac 1r \right) \, , \;\;\;\;
{\ssr Q}_{3A||B}\varepsilon^{AB}=O\left(\frac 1r \right) \]
\noindent
On each sphere
$S(r)$ the full information about tensor $Q_{\mu\nu}$ is encoded in the
following 6 scalar functions: $Q_{03}$, $Q_{0A||B}\eta^{AB}$,
$Q_{3A||B}\eta^{AB}$, $Q_{0A||B}\varepsilon^{AB}$,
$Q_{0A||B}\varepsilon^{AB}$, $q$ where $q:=\frac 12
Q_{AB}\varepsilon^{AB}$. The full set of the above components splits, in a
natural way, into two sets:
$\{$ $Q_{03}$, $Q_{0A||B}\eta^{AB}$,
$Q_{3A||B}\eta^{AB}$ $\}$ and $\{$ $Q_{0A||B}\varepsilon^{AB}$,
$Q_{0A||B}\varepsilon^{AB}$, $q$ $\}$. Let us notice that the equations
(\ref{003}--\ref{ABC}) can be also splitted in the same way.
The ``dynamics'' of the first set can be described by the following
relations:
\be\label{003p}
3{\cal Q}_{003}=3v{\cal Q}_{3AB}\eta^{AB}=
-2\dot{Q}_{03} + v Q_{3A||B}\eta^{AB}
\ee

\be\label{00Ap}
3r{\cal Q}_{00}{^A}{_{||A}}=
-2r\dot{Q}_0{^A}{_{||A}} +  \frac{2mv}{r}  Q_{3}{^A}{_{||A}}
- v^2 r Q_{3A||B,3}\eta^{AB}
\ee

\be\label{03Ap}
r{\cal Q}_{03}{^A}{_{||A}}= -r\dot{Q}_{3}{^A}{_{||A}} +
r{Q}_{03}{^{||A}}{_{||A}} -  Q_{0}{^A}{_{||A}}
 + \frac m{vr} Q_{0}{^A}{_{||A}}
\ee

\be\label{033p}
3{\cal Q}_{033}=-\frac 3v{\cal Q}_{0AB}\eta^{AB}=
 \frac3{2v}{\cal Q}_{A0B}\eta^{AB} =
2{Q}_{03,3} -  \frac{2m}{vr^2} Q_{03} -  \frac{2}{r} Q_{03}
- \frac1{v} Q_{0}{^A}{_{||A}}
\ee

\be\label{30Ap}
r{\cal Q}_{30}{^A}{_{||A}}=  -r\left(Q_{0}{^A}{_{||A}}\right)_{,3}
-r{Q}_{03}{^{||A}}{_{||A}}  + \frac m{vr} Q_{0}{^A}{_{||A}}
\ee

\be\label{00A-33A}
\frac r{v^2}{\cal Q}_{00}{^A}{_{||A}}-2r{\cal Q}_{33}{^A}{_{||A}}=
r\left( {Q}_{3}{^A}{_{||A}} \right)_{,3}
\ee
There are more equations but they are linearily dependent, for example:
\[
3v^{-1}r{\cal Q}_{00}{^A}{_{||A}}
-3vr {\cal Q}_{33}{^A}{_{||A}} =3r{\cal Q}^{ABC}{_{||A}}\eta_{BC}=
 -\frac rv \dot{Q}_{0}{^A}{_{||A}}  - v Q_{3}{^A}{_{||A}} +
 \]
  \be\label{ABCp}
  + \frac{m}{r}  Q_{3}{^A}{_{||A}}
  +v r\left( Q_{3}{^A}{_{||A}} \right)_{,3}
\ee
{}From (\ref{003p}) and (\ref{033p}) we obtain the following exact solution
for the monopole part of $Q_{03}$:
\[ \ssm Q_{03}=rv^{-\frac12} \]
\noindent
The dipole part $\ssd Q_{03}$, $\ssd Q_{0A||B}\eta^{AB}$,
$\ssd Q_{3A||B}\eta^{AB}$ can be obtained from eq.
(\ref{003}--\ref{00A-33A}) only asymptotically:

\[ 2\ssd Q_{03} = K \left( r^2-mr - t^2 \right) + P t \]

\be\label{KP}
 r\ssd Q_0{^A}{_{||A}} = K \left( r^2-3mr + t^2 \right) -Pt \ee

\[ r\ssd Q_3{^A}{_{||A}} = Kt ( 5m-2r) +Pr \]
\noindent
where $K$, $P$ are dipole functions on a unit sphere and solution is
unique up to $O(1)$.

Similarly for $Q_{3A||B}\varepsilon^{AB}$,
$Q_{0A||B}\varepsilon^{AB}$ and $q$ we have the following relations:

\be\label{3Ad}
r{\cal Q}_{03A||B}\varepsilon^{AB}= -r\dot{Q}_{3A||B}\varepsilon^{AB}
 -  Q_{0A||B}\varepsilon^{AB}
 + \frac m{vr} Q_{0A||B}\varepsilon^{AB}
\ee

\be\label{ABd}
{\cal Q}_{0AB}\varepsilon^{AB}=
 Q_{0A||B}\varepsilon^{AB} - 2\dot{q}
\ee

\[
 \frac r{v^2}{\cal Q}_{00A||B}\varepsilon^{AB}
 +r{\cal Q}_{33A||B}\varepsilon^{AB}=
  -\frac r{v^2}\dot Q_{0A||B}\varepsilon^{AB}
 -\left( r{Q}_{3A||B}\varepsilon^{AB} \right)_{,3} +\]
 \be\label{0Ad}
   + \frac m{vr}Q_{3A||B}\varepsilon^{AB}
\ee

\be\label{3A-aq}
 \frac r{v^2}{\cal Q}_{00A||B}\varepsilon^{AB}
 -2r{\cal Q}_{33A||B}\varepsilon^{AB}=
r\left( {Q}_{3A||B}\varepsilon^{AB} \right)_{,3} +\frac 1{rv} {\bf a}q
\ee

\be\label{3A-q3}
{\cal Q}_{3AB}\varepsilon^{AB}=Q_{3A||B}\varepsilon^{AB} -2q_{,3}
 +\frac 2r q
\ee

\be\label{0A}
  r{\cal Q}_{30A||B}\varepsilon^{AB}=
 -r\left(Q_{0A||B}\varepsilon^{AB}\right)_{,3}
  + \frac m{vr} Q_{0A||B}\varepsilon^{AB}
\ee
We obtain an exact monopole solution from (\ref{ABd}) and (\ref{3A-q3}):
\[ \ssm q=r \]
\noindent
An asymptotic solution of (\ref{3Ad}--\ref{0A})
 for $\ssd Q_{3A||B}\varepsilon^{AB}$,
  $\ssd Q_{0A||B}\varepsilon^{AB}$ and $\ssd q$ has the following form (up
to $O(1)$):
\samepage{\[ 2\ssd q = J \left( r^2-2mr -t^2 \right) +Bt \]
 \be\label{JB}
  r\ssd Q_{3A||B}\varepsilon^{AB} = J\left( r^2+t^2 \right) -Bt \ee
\[ r\ssd Q_{0A||B}\varepsilon^{AB} = Jt( 9m-2r) +Br \]}
\noindent
where again $J$ and $B$ are dipoles on a unit sphere:
\[ ({\bf a} +2)J =({\bf a} +2)B=J_{,3}=B_{,3}=\dot J =\dot B=0 \]

It is easy to check that for $m=0$ the dipole solutions (\ref{KP}) and
(\ref{JB}) are exact.
This way we get 14 dimensional space of CYK tensors for flat Minkowski
space  (12 dipoles plus 2 monopoles). In this case we have also 6 more
constant CYK tensors fullfilling equation (\ref{CYK}) so the total space
of CYK tensors for flat Minkowski space is 20 dimensional \cite{JJ-s2},
\cite{Pen-Rin}.

\end{document}